\documentclass[twocolumn,pra,aps]{revtex4} 
\usepackage{epsfig,amsmath}
\usepackage{graphicx}
\usepackage{dcolumn}
\usepackage{bm}
\usepackage{bbold}
\newcommand{\beq}{\begin{equation}}
\newcommand{\eeq}{\end{equation}}
\newcommand{\beqa}{\begin{eqnarray}}
\newcommand{\eeqa}{\end{eqnarray}}

\begin{document}
\author{Yi Ding}
\email{yding2020swjtu@163.com}
\affiliation{School of Physical Science and Technology, Southwest Jiaotong University, Chengdu 610031, China}
\title{Hanbury Brown-Twiss effect with electromagnetic scattered field generated by a collection of particles of $\mathcal{L}$ types}
\date{\today}

\begin{abstract}
A theoretical framework in the spherical polar coordinate system is developed to systematically treat the correlation between intensity fluctuations (CIF) of electromagnetic light waves on scattering from a collection of particles of $\mathcal{L}$ types. Two $\mathcal{L}\times \mathcal{L}$ matrices called pair-potential matrix (PPM) and pair-structure matrix (PSM) are introduced to jointly formulate the normalized CIF of the scattered field for the first time. We build a closed-form relation that associates the normalized CIF with the PPM and the PSM as well as the spectral degree of polarization $\mathcal{P}$ of the incident field, showing that the normalized CIF is closely related to the trace of the product of the PSM and the transpose of the PPM, and its dependence on $\mathcal{P}$ is completely determined by the scattering polar angle and azimuth angle. For a special case where the spatial distributions of scattering potentials of particles of different types are similar and the same is true of their density distributions, the PPM and the PSM will reduce to two new matrices whose elements separately quantify the degree of angular correlation of the scattering potentials of particles and their density distributions, and the number of species of particles in this special case appears as a scaled factor to ensure the normalization of the CIF. The effects of the off-diagonal elements of the PPM and the PSM on the normalized CIF and its dependence on $\mathcal{P}$ are illustrated by two numerical examples.      

\end{abstract}

\maketitle

\section{Introduction}

It is well-known that the Hanbury Brown-Twiss effect \cite{HBT} is a manifestation of the correlation between intensity fluctuations (CIF) at two observation points in an optical field. The traditional analysis of the HBT effect omits the polarization properties of light fields and thus a scalar treatment is usually enough. However, with the extension of CIF to the domain of random electromagnetic fields, it has been shown that the normalized CIF is closely related to the degree of polarization of light fields via an elegant mathematical identity \cite{CE, SLK}, for electromagnetic fields obeying Gaussian statistics. The further studies have shown that the knowledge of the degree of polarization of light fields is not adequate to determine the CIF, and the notion of degree of cross-polarization needs to be introduced \cite{SW, SJS, ALK}, but the usefulness of this notion was later questioned \cite{HTS}. Additionally, the propagation dynamics of the CIF of electromagnetic Gaussian Schell-model beams in free space \cite{WV} and in atmospheric turbulence \cite{OK1} has been addressed, and the generalized Hanbury Brown–Twiss effect in partially coherent electromagnetic beams have also been executed \cite{WDV}.     

A novel application of the Hanbury Brown-Twiss effect to weak scattering theory can be traced back to the study of the CIF of a scalar plane wave on scattering from a quasi-homogeneous random medium \cite{Xin}. Such a pioneering work shed light on the relation between the CIF of the scattered field and the structural characteristics of the medium, and laid a foundation of the application of CIF to inverse scattering problem \cite{DVW, LS}. Shortly afterwards, this investigation was extended to the scattering of stochastic scalar fields \cite{JK} and electromagnetic light waves \cite{Li}. It is found that in addition to the structural characteristics of the medium, the coherence and polarization properties of incoming electromagnetic sources are also important elements for determining the CIF in the far-zone scattered field. In recent years, the CIF in random electromagnetic scattered fields continues to be of great interest, among them \cite{ZZ, LCC, LCC1, Wang, Ding}, in which the CIF of electromagnetic light waves on scattering has been extended to a system of identical particles. However, as we know, a particulate system in which all particles are identical is actually a pretty idealized model, and it has no ability to characterize a more practical situation where particles in the collection have different sizes and shapes, as well as various distribution features of refractive index. Such a particulate collection is often encountered in practice, unfortunately, the CIF of electromagnetic light waves on scattering from a collection of particles of different types hasn't been addressed. Additionally, although there have been many efforts to explore the influence of the polarization property of the incident field on the normalized CIF of the scattered field, the qualitative and quantitative mathematical relation between the spectral degree of polarization $\mathcal{P}$ of the incident field and the normalized CIF of the scattered field hasn't been fully exposed.  

In this work, we will develop a theoretical framework in the spherical polar coordinate system to systematically treat the CIF of electromagnetic light waves on scattering from a collection of particles of $\mathcal{L}$ types. Two $\mathcal{L}\times \mathcal{L}$ matrices called pair-potential matrix (PPM) and pair-structure matrix (PSM) will be introduced to jointly formulate the normalized CIF of the scattered field. We will derive an analytic expression that elaborates the qualitative and quantitative relation between the normalized CIF of the scattered field and the PPM, the PSM, the spectral degree of polarization $\mathcal{P}$ of the incident field. The influence of the off-diagonal elements of the PPM and the PSM on the normalized CIF of the scattered field and its dependence on $\mathcal{P}$ will be discussed in detail by analyzing the scattering of electromagnetic light waves from two special hybrid particulate systems, i.e., a collection of random particles with determinate density distributions and a collection of determinate particles with random density distributions.

\section{Convolution representation of the scattering potential and its correlation function of a collection of particles with $\mathcal{L}$ types}

For a collection of particles, there are usually $\mathcal{L}$ types of particles forming this system, $m(p)$ of each type, ($p=1,2,3,\cdot\cdot\cdot,\mathcal{L}$), located at points specified by position vectors $\mathbf{r}_{pm}$. We characterize the response of each particle to an incoming field by a scattering potential $f_{p}(\mathbf{r}^{\prime})$, which is closely related to the refractive index of the particle (\cite{Wolf2}, Sec. 6.1). The scattering potential $F(\mathbf{r}^{\prime},\omega)$ of the whole collection can be usually defined as \cite{DW} \begin{align}\label{scatteringpotential}
    F(\mathbf{r}^{\prime},\omega)=\sum_{p=1}^{\mathcal{L}}\sum_{m(p)}f_{p}(\mathbf{r}^{\prime}-\mathbf{r}_{pm},\omega)
\end{align}
For the sake of following discussions, we now rewrite the definition of the scattering potential of the collection in a slightly unfamiliar form, viz., 
\begin{equation}\label{scatteringpotential1}
    F(\mathbf{r}^{\prime},\omega)=\sum_{p=1}^{\mathcal{\mathcal{L}}}f_{p}(\mathbf{r}^{\prime},\omega)\otimes g_{p}(\mathbf{r}^{\prime}),
\end{equation}
where 
\begin{equation}\label{densityfunction}
    g_{p}(\mathbf{r}^{\prime})\equiv\sum_{m(p)}\delta(\mathbf{r}^{\prime}-\mathbf{r}_{pm})
\end{equation}
may be interpreted as the density function of the $p$th-type particle \cite{Gbur} if the $p$th-type particle in the collection can be effectively regarded as consisting of $m(p)$ `point particles'. $\delta(\cdots)$ is the three-dimensional Dirac delta function, and $\otimes$ denotes the convolution operation.

In general, the collection may be of deterministic or random nature. In the case when the collection is deterministic, its scattering potential $F(\mathbf{r}^{\prime},\omega)$ is a well-defined function of position. However, for a more involved case when the scattering potential of the collection is not deterministic, but varies randomly as a function of position. In this case, the spatial correlation function of scattering potential at two points $\mathbf{r_{1}}^{\prime}$ and $\mathbf{r_{2}}^{\prime}$ may be given as (\cite{Wolf2}, Sec. 6.3.1)
\begin{equation}\label{correlationfunction}
{C_{F}}\left( {{{\mathbf{{r_{1}^{\prime}}}}},{{\mathbf{{r_{2}^{\prime}}}}}%
,\omega}\right) =\left\langle {{F^{\ast}}\left( {{{\mathbf{{r_{1}^{\prime}%
}}}},\omega}\right) F\left( {{{\mathbf{{r_{2}^{\prime}}}}},\omega}\right)
}\right\rangle_{M},
\end{equation}
where $\left\langle\cdots\right\rangle_{M}$ stands for the average taken over different realizations of the scatterer. On substituting from Eq. \eqref{scatteringpotential1} into Eq. \eqref{correlationfunction}, after some simple rearrangements, we end up with
\begin{equation}\label{sc}
 {C_{F}}\left( {{{\mathbf{{r_{1}^{\prime}}}}},{{\mathbf{{r_{2}^{\prime}}}}}%
,\omega}\right)=\sum_{p=1}^{\mathcal{L}}\sum_{q=1}^{\mathcal{L}}{C_{f_{pq}}}\left({{{\mathbf{{r_{1}^{\prime}}}}},{{\mathbf{{r_{2}^{\prime}}}}}%
,\omega}\right)\otimes{C_{g_{pq}}}\left({{{\mathbf{{r_{1}^{\prime}}}}},{{\mathbf{{r_{2}^{\prime}}}}}%
}\right).
\end{equation}
where 
\begin{equation}\label{scatteringcorrelation}
   {C_{f_{pq}}}\left({{{\mathbf{{r_{1}^{\prime}}}}},{{\mathbf{{r_{2}^{\prime}}}}},\omega}\right)=\left\langle f_{p}^{*}(\mathbf{r}_{1}^{\prime},\omega)f_{q}(\mathbf{r}_{2}^{\prime},\omega)\right\rangle 
\end{equation}
represents the self-correlation functions of the scattering potentials of particles of same type (if $p=q$) or the cross-correlation functions of the scattering potentials of particles of different types (if $p\neq q$), and 
\begin{equation} \label{densitycorrelation}
{C_{g_{pq}}}\left({{{\mathbf{{r_{1}^{\prime}}}}},{{\mathbf{{r_{2}^{\prime}}}}}}\right)=\left\langle g_{p}^{*}(\mathbf{r}_{1}^{\prime})g_{q}(\mathbf{r}_{2}^{\prime})\right\rangle    
\end{equation}
represents the self-correlation functions of the density distributions of particles of same type (if $p=q$) or the cross-correlation functions of the density distributions of particles of different types (if $p\neq q$).

It is seen in the transition from Eqs. \eqref{scatteringpotential1} and \eqref{correlationfunction} to Eq. \eqref{sc} that use has been made of the assumption that the average over the ensemble of the scattering potentials of particles and that over the ensemble of their density distributions are mutually independent. 

\section{Correlation between intensity fluctuations of electromagnetic light waves
on scattering from a collection with particles of $\mathcal{L}$ types in the spherical polar coordinate system}

\begin{figure}[bthp]
\centering
\includegraphics[width=8cm]{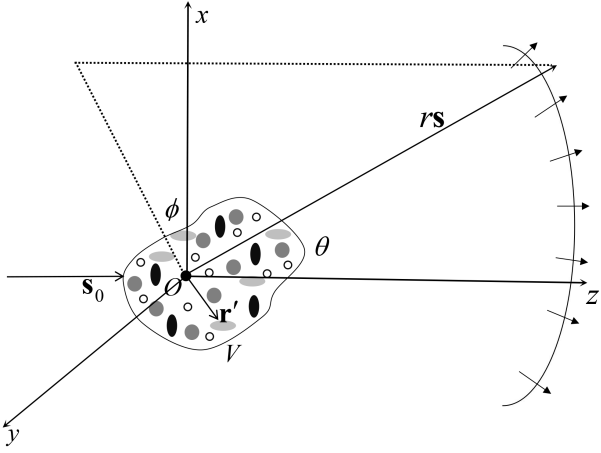}
\caption{Illustration of notations.}
\label{Fig 1}
\end{figure}

Assume now an electromagnetic plane wave, propagating in the direction of a unit vector $\mathbf{s}_{0}$ along the $z$ axis, is incident on a collection of particles with $\mathcal{L}$ types (see Fig. \ref{Fig 1}). The incident field at a point $\mathbf{r}^{\prime}$ can be characterized by a statistical ensemble $\left\{E_{i}(\mathbf{r}^{\prime},\omega)\exp{[-i\omega t}]\right\}$ of monochromatic realizations, all of frequency $\omega$, in the sense of coherence theory in the space-frequency domain. Here
\begin{align}\label{infield}
E_{i}(\mathbf{r}^{\prime},\omega)=A_{i}a_{i}(\omega)\exp(ik\mathbf{s}_{0} \cdot \mathbf{r}^{\prime}), \ (i=x,y),
\end{align}
where $A_{i}$ is a constant, representing the amplitude of the electric field along the $i$th axis, $k = \omega/c$ is the wave number with $c$ being the speed of light in vacuum, and $a_{i}(\omega)$ is (generally complex) frequency-dependent random variables. Here, we assume $a_{x}(\omega)=a_{y}(\omega)=a(\omega)$. 

The cross-spectral density matrix of the incident field at a pair of points $\mathbf{r_{1}}^{\prime}$ and $\mathbf{r_{2}}^{\prime}$ may be expressed in the form (\cite{Wolf2}, Sec. 9.1)
\begin{align}\label{CSDM}
{\mathbf{W}^{(\mathrm{i})}}({{{\mathbf{{r_{1}^{\prime }}}}},{{\mathbf{{%
r_{2}^{\prime }}}}}}{,\omega })& \equiv \left[ {{W_{ij}^{(%
\mathrm{i})}}({{\mathbf{{r_{1}^{\prime }}}}},{{\mathbf{{r_{2}^{\prime }}}}}}{,\omega )}\right]  \notag \\
& =\left[ {\left\langle {{E_{i}}{^{\ast }}\left( {{{\mathbf{{r_{1}^{\prime }}%
}}},\omega }\right) {E_{j}}\left( {{{\mathbf{{r_{2}^{\prime }}}}},\omega }%
\right) }\right\rangle }\right]  \notag \\
& \;\left( {i=x,y;\ j=x,y}\right) ,
\end{align}%
where the asterisk denotes the complex conjugate and the angular brackets denote ensemble average. On substituting from Eq. (\ref{infield}) into Eq. (\ref{CSDM}), it follows that the elements of the cross-spectral density matrix of the incident field can be calculated as
\begin{align}\label{element}
  W_{ij}^{(\text{i})}(\mathbf{r_{1}}^{\prime},\mathbf{r_{2}}^{\prime},\omega)=A_{i}A_{j}B_{ij}S(\omega)\exp[ik\mathbf{s_{0}} \cdot (\mathbf{r}_{1}^{\prime}-\mathbf{r}_{2}^{\prime})],
\end{align}
where
\begin{equation}\label{spectrum}
S(\omega)=\left\langle a^{*}(\omega)a(\omega)\right\rangle
\end{equation}
is the spectrum of the incident field. For simplicity, we assume that the $x$ and $y$ components of the electric field falling on the scatterer are uncorrelated (\cite{Wolf2}, Sec. 9.1; \cite{SW1}), i.e., $B_{ij}=1$ if and only if $i=j$, otherwise $B_{ij}=0$.

Clearly, the cross-spectral density matrix of the incident field is now diagonal, and the spectral degree of polarization $\mathcal{P}$ of the incident field in this situation has a simple form (\cite{Wolf2}, Sec. 9.2)
\begin{equation}\label{polarization}
    \mathcal{P}\equiv\Bigg[1-\frac{4\text{Det}\mathbf{W}^{\text{(i)}}(\mathbf{r}^{\prime}_{1},\mathbf{r}^{\prime}_{2},\omega)}{\bigl[\text{Tr}\mathbf{W}^{\text{(i)}}(\mathbf{r}^{\prime}_{1},\mathbf{r}^{\prime}_{2},\omega)\bigr]^{2}}\Biggr]^{\frac{1}{2}}=\Biggl|\frac{A_{x}^{2}-A_{y}^{2}}{A_{x}^{2}+A_{y}^{2}}\Biggr|,
\end{equation}
where $\text{Det}$ and $\text{Tr}$ denote the determinant and the trace, respectively. When $\mathcal{P}=1$ the incident field is, in this case, said to be completely polarized across the source. In the other extreme case, when $\mathcal{P}=0$ the incident field is said to be completely unpolarized across the source. In the intermediate case, when $0<\mathcal{P}<1$, the incident field is said to be partially polarized across the source.

It is well known that within the validity of the first-order Born approximation, the electric vector of the scattered field in the far zone out of scatterer can be formulated as \cite{BW}
\begin{align}\label{electricvector}
    \mathbf{E}^{(\text{s})}(r\mathbf{s},\omega)&=-\frac{e^{ikr}}{r}\mathbf{s}\times \Bigl[\mathbf{s}\times\int_{\mathcal{V}}[\sum_{i=1}^{\mathcal{L}}f_{i}(\mathbf{r}^{\prime},\omega)\otimes  g_{i}(\mathbf{r}^{\prime})] \notag \\ & \times \mathbf{E}^{\text{(i)}}(\mathbf{r}^{\prime},\omega)e^{-ik\mathbf{s}\cdot \mathbf{r}^{\prime}}d^{3}r^{\prime}\Bigr],
\end{align}
where $\mathbf{s}$ is the direction of the scattering path and $\mathbf{E}^{\text{(i)}}(\mathbf{r}^{\prime},\omega)$ is the electric vector of the incident field.

From Eq. (\ref{electricvector}), one can notice that $\mathbf{s}\cdot\mathbf{E}^{(\text{s})}(r\mathbf{s},\omega)=0$, i.e., only two transverse components of the scattered field are nontrivial, and the overall behavior of the scattered field in the far zone is essentially of an outgoing spherical wave. In this sense, the scattered field can be greatly simplified if one expresses it in the spherical polar coordinate system \cite{Tong1}, viz.,
\begin{subequations}\label{polarcoordinate}
\begin{align}
    E_{\theta}^{\text{(s)}}(r\mathbf{s},\omega)&=-\frac{e^{ikr}}{r}\int_{\mathcal{V}}[\sum_{i=1}^{\mathcal{L}}f_{i}(\mathbf{r}^{\prime},\omega)\otimes g_{i}(\mathbf{r}^{\prime})][\cos{\theta}\cos{\phi} \notag \\ & \times E_{x}^{\text{(i)}}(\mathbf{r}^{\prime},\omega)  +\cos{\theta}\sin{\phi}E_{y}^{\text{(i)}}(\mathbf{r}^{\prime},\omega)]\notag \\ & \times e^{-ik\mathbf{s}\cdot \mathbf{r}^{\prime}} d^{3}r^{\prime}, \\
    E_{\phi}^{\text{(s)}}(r\mathbf{s},,\omega)&=-\frac{e^{ikr}}{r}\int_{\mathcal{V}}[\sum_{i=1}^{\mathcal{L}}f_{i}(\mathbf{r}^{\prime},\omega)\otimes g_{i}(\mathbf{r}^{\prime})] \notag \\ & \times [-\sin{\theta}E_{x}^{\text{(i)}}(\mathbf{r}^{\prime},\omega)+\cos{\phi}E_{y}^{\text{(i)}}(\mathbf{r}^{\prime},\omega)]
    \notag \\ & \times e^{-ik\mathbf{s}\cdot \mathbf{r}^{\prime}}d^{3}r^{\prime}.
\end{align}
\end{subequations}

The second-order coherence and polarization properties of the scattered
field at two points $r\mathbf{s}_{1}$ and $r\mathbf{s}_{2}$ can also be
characterized by a $2\times2$ cross-spectral density matrix, with a form of (\cite{Wolf2}, Sec. 9.1)
\begin{align}\label{Scrossdensity}
{\mathbf{W}^{(\text{s})}}(r{\mathbf{s}_{1}},r{\mathbf{s}_{2}},\omega )& \equiv %
\left[ {{W_{ij}^{(\text{s})}}(r{\mathbf{s}}}_{1}{,r{\mathbf{s}}}_{2}{%
,\omega )}\right]   \notag \\
& =\left[ {\left\langle {{E_{i}^{(\text{s})}}{^{\ast }}\left( {r{{%
\mathbf{s}}}}_{1}{,\omega }\right) {E_{j}^{(\text{s})}}\left( {r{%
\mathbf{s}}}_{2}{,\omega }\right) }\right\rangle }\right]   \notag \\
& \;\left( {i=\theta,\phi;\ j=\theta,\phi}\right).
\end{align}%

We now consider the intensity of a single realization of the scattered field at a point $r\mathbf{s}$ at frequency $\omega$, which can be expressed as
\begin{align}\label{intensity1}
{I^{(\text{s})}}\left( {r\mathbf{s},\omega}\right) & =\Bigl|{E_{\theta}^{(\text{s})}}\left( {r\mathbf{s},\omega}\right)\Bigr|^{2}+\Bigl|{
E_{\phi}^{(\text{s})}}\left( {r\mathbf{s},\omega}\right)\Bigr|^{2}.
\end{align}%

The intensity dispersion from its mean value at the same point $r\mathbf{s}$ is given as
\begin{equation}\label{deltaintensity}
\Delta{I^{(\text{s})}}\left( {r\mathbf{s},\omega}\right) ={I^{\left(
s\right) }}\left( {r\mathbf{s},\omega}\right) -\left\langle {{I^{\left(
s\right) }}\left( {r\mathbf{s},\omega}\right) }\right\rangle ,
\end{equation}
where $\left\langle {{I^{(\text{s})}}\left( {r\mathbf{s},\omega}\right) }
\right\rangle=\mathrm{Tr}{\mathbf{W}^{(\text{s})}}(r{\mathbf{s}_{1}},r{\mathbf{s}_{2}},\omega)$. 

We adopt the HBT coefficient as a measure of the correlation of intensity fluctuations at two points ${r\mathbf{s}}_{1}$\ and ${r\mathbf{s}}_{2}$, which is defined as
\begin{equation}\label{definition}
{\mathcal{C}}\left( {r{\mathbf{s}_{1}},r{\mathbf{s}_{2}},\omega}%
\right) \equiv\left\langle {\Delta{I^{(\text{s})}}\left( {r{\mathbf{s}%
_{1}},\omega}\right) \Delta{I^{(\text{s})}}\left( {r{\mathbf{s}_{2}}%
,\omega}\right) }\right\rangle.
\end{equation}

If we assume that the fluctuations of the scattered field obey the Gaussian
statistics as is often the case, and this allows the calculation of the
fourth-order correlation from the second-order one (\cite{Wolf2}, Sec. 7.2). In this way, the
CIF of the scattered field can be simplified as
\begin{equation}\label{CIF}
  {\mathcal{C}}\left( {r{\mathbf{s}}}_{1}{,r{\mathbf{s}_{2}},\omega}%
\right)=\text{Tr}[{\mathbf{W^{\dag}}^{(\text{s})}}(r{\mathbf{s}_{1}},r{\mathbf{s}_{2}},\omega )\cdot{\mathbf{W}^{(\text{s})}}(r{\mathbf{s}_{1}},r{\mathbf{s}_{2}},\omega )],
\end{equation}
where $\cdot$ and $\dag$ stand for the ordinary multiplication operation and Hermitian adjoint, respectively. 

In many situations of practical interest, one often takes into account the normalized version of the CIF, which can be defined in terms of Eq. (\ref{CIF}) as the following form (\cite{Wolf2}, Sec. 7.2)
\begin{align}\label{NCIF}
{\mathcal{C}_{n}}\left( {r{\mathbf{s}}}_{1}{,r{\mathbf{s}_{2}},\omega}%
\right)&\equiv \frac{{\mathcal{C}}\left( {r{\mathbf{s}_{1}},r{\mathbf{s}_{2}},\omega}%
\right)}{\left\langle {{I^{\left(
\text{s}\right) }}\left( {r\mathbf{s}_{1},\omega}\right) }\right\rangle\left\langle {{I^{\left(
\text{s}\right) }}\left( {r\mathbf{s}_{2},\omega}\right)  }\right\rangle}\notag \\ &=\frac{\text{Tr}[{\mathbf{W^{\dag}}^{(\text{s})}}(r{\mathbf{s}_{1}},r{\mathbf{s}_{2}},\omega )\cdot{\mathbf{W}^{(\text{s})}}(r{\mathbf{s}_{1}},r{\mathbf{s}_{2}},\omega )]}{\mathrm{Tr}{\mathbf{W}^{(\text{s})}}(r{\mathbf{s}_{1}},r{\mathbf{s}_{1}},\omega)\mathrm{Tr}{\mathbf{W}^{(\text{s})}}(r{\mathbf{s}_{2}},r{\mathbf{s}_{2}},\omega)}.
\end{align}

On substituting from Eqs. (\ref{infield}) and (\ref{polarcoordinate}) first into (\ref{Scrossdensity}), and then inserting the results into Eq. (\ref{NCIF}), after some long but straightforward calculation, we end up with
\begin{small}
\begin{align}\label{NCIF11}
 {\mathcal{C}_{n}}\left( {r{\mathbf{s}}}_{1}{,r{\mathbf{s}_{2}},\omega}%
\right)&=\frac{\alpha_{1}\alpha_{2}A_{x}^{4}+\beta_{1}\beta_{2}A_{x}^2A_{y}^2+\varepsilon_{1}\varepsilon_{2}A_{y}^4}{\alpha_{1}\alpha_{2}A_{x}^{4}+(\alpha_{1}\varepsilon_{2}+\alpha_{2}\varepsilon_{1})A_{x}^{2}A_{y}^{2}+\varepsilon_{1}\varepsilon_{2}A_{y}^{4}}\notag \\ &\times \Biggl|\sum_{p,q}^{\mathcal{L}}\widetilde{C}_{f_{pq}}(-\mathbf{K}_{1},\mathbf{K}_{2},\omega)\widetilde{C}_{g_{pq}}(-\mathbf{K}_{1},\mathbf{K}_{2},\omega)\Biggr|^2 \notag \\ &\times \Biggl(\sum_{p,q}^{\mathcal{L}}\widetilde{C}_{f_{pq}}(-\mathbf{K}_{1},\mathbf{K}_{1},\omega)\widetilde{C}_{g_{pq}}(-\mathbf{K}_{1},\mathbf{K}_{1},\omega)\notag \\ &\times\sum_{p,q}^{\mathcal{L} }\widetilde{C}_{f_{pq}}(-\mathbf{K}_{2},\mathbf{K}_{2},\omega)\widetilde{C}_{g_{pq}}(-\mathbf{K}_{2},\mathbf{K}_{2},\omega)\Biggr)^{-1}
\end{align}
\end{small}
where
\begin{align}\label{parameters}
    \alpha_{i} &=\cos^{2}({\theta_{i}})\cos^{2}({\phi_{i}})+\sin^{2}({\phi_{i}})  \notag \\ 
    \beta_{i} &=\frac{\sqrt{2}}{4}\sin({2\phi_{i}})(1-\cos({2\theta_{i}})) \qquad (i=1,2) \\ 
    \varepsilon_{i} &=\cos^{2}({\theta_{i}})\sin^{2}({\phi_{i}})+\cos^{2}({\phi_{i}}), \notag
\end{align}
and
\begin{align}\label{1}
  \widetilde{C}_{f_{pq}}(\mathbf{K}_{1},\mathbf{K}_{2},\omega)&=\int_{\mathcal{V}}\int_{\mathcal{V}}{C_{f_{pq}}}\left({{{\mathbf{{r_{1}^{\prime}}}}},{{\mathbf{{r_{2}^{\prime}}}}}%
,\omega}\right) \notag \\ & \times \exp{\bigl[-i(\mathbf{K}_{2}\cdot{r}_{2}^{\prime}+\mathbf{K}_{1}\cdot{r}_{1}^{\prime})\bigr]}d^{3}r_{1}^{\prime}d^{3}r_{2}^{\prime}
\end{align}stands for the six-dimensional spatial Fourier transformations of the self-correlation functions of the scattering potentials of particles of the same type (if $p=q$) or the cross-correlation functions of the scattering potentials of particles of different types (if $p\neq q$), with $\mathbf{K_{1}}=k(\mathbf{s}_{1}-\mathbf{s}_{0})$ and $\mathbf{K_{2}}=k(\mathbf{s}_{2}-\mathbf{s}_{0})$ being analogous to the momentum transfer vector of quantum mechanical theory of potential scattering, and 
\begin{align}\label{2}
  \widetilde{C}_{g_{pq}}(\mathbf{K}_{1},\mathbf{K}_{2},\omega)&=\int_{\mathcal{V}}\int_{\mathcal{V}}{C_{g_{pq}}}\left({{{\mathbf{{r_{1}^{\prime}}}}},{{\mathbf{{r_{2}^{\prime}}}}}}\right)\notag \\ & \times \exp{\bigl[-i(\mathbf{K}_{2}\cdot{r}_{2}^{\prime}+\mathbf{K}_{1}\cdot{r}_{1}^{\prime})\bigr]}d^{3}r_{1}^{\prime}d^{3}r_{2}^{\prime}
\end{align}stands for the six-dimensional spatial Fourier transformations of the self-correlation functions of the density distributions of particles of the same type (if $p=q$) or the cross-correlation functions of the density distributions of particles of different types (if $p\neq q$).

\section{Relation between the normalized CIF of the scattered field and the PPM, the PSM, the spectral degree of polarization $\mathcal{P}$ of the incident field}
We now introduce two $\mathcal{L}\times \mathcal{L}$ matrices to jointly formulate the normalized CIF of the scattered field. These two matrices not only provide sufficient amount of information to determine all the second-order statistical properties of the scattered field, but also can largely simplify the theoretical procedures of the scattering of light from complex collection of scatters. The first matrix is defined as 
\begin{align}\label{matrix1}
    \mathcal{F}(\mathbf{K_{1}},\mathbf{K_{2}},\omega)&=\Bigl[\widetilde{C}_{f_{pq}}(-\mathbf{K_{1}},\mathbf{K_{2}},\omega)\Bigr]_{\mathcal{L}\times\mathcal{L}}.
\end{align}From Eqs. \eqref{scatteringcorrelation} and \eqref{1}, it readily follows that the diagonal elements of this matrix represent angular the self-correlations of the scattering potentials of particles of same type, while the off-diagonal elements represent the angular cross-correlations of the scattering potentials of each pair of particle types. The entire matrix contains all the information about the angular correlation properties of the scattering potentials of between particles within one type and across different types, it may be natural to call it the pair-potential matrix (PPM) \cite{DDY}, which hasn't been noticed before.

The second matrix is defined as
\begin{align}\label{matrix2}
    \mathcal{G}(\mathbf{K_{1}},\mathbf{K_{2}},\omega)&=\Bigl[\widetilde{C}_{g_{pq}}(-\mathbf{K_{1}},\mathbf{K_{2}},\omega)\Bigr]_{\mathcal{L}\times\mathcal{L}}.
\end{align}From Eqs. \eqref{densitycorrelation} and \eqref{2}, it also readily accessible that the diagonal elements of this matrix stand for the angular self-correlations of the density distributions of particles of same type, while the off-diagonal elements stand for the angular cross-correlations of the density distributions of each pair of particle types. The entire matrix contains all the information about the angular correlation properties of the density distributions of between particles within one type and across different types. Notice that $\mathcal{G}(\mathbf{K_{1}},\mathbf{K_{2}},\omega)$ is essentially the same as the pair-structure matrix (PSM) introduced by Tong $\textit{et al}$ before \cite{Tong}, and one can see from here that the pair-structure factors and joint pair-structure factors are related to the angular correlations of the density distributions of particles within one type and across different types in the collection, respectively.

One may also need to notice that $\mathcal{F}(\mathbf{K_{1}},\mathbf{K_{2}},\omega)$ and $\mathcal{G}(\mathbf{K_{1}},\mathbf{K_{2}},\omega)$ are, in general, not Hermitian matrices since $\widetilde{C}_{{f}_{qp}}(-\mathbf{K_{2}},\mathbf{K_{1}},\omega)\neq\widetilde{C}_{f_{pq}}^{*}(-\mathbf{K_{1}},\mathbf{K_{2}},\omega)$ and $\widetilde{C}_{g_{qp}}(-\mathbf{K_{2}},\mathbf{K_{1}},\omega)\neq\widetilde{C}_{g_{pq}}^{*}(-\mathbf{K_{1}},\mathbf{K_{2}},\omega)$. However, in many situations of practical interest, i.e., their elements have common Gaussian Schell-model distributions \cite{Dingz} and multi-Gaussian Schell-model distributions \cite{ZJZ} as well as quasi-homogeneous distributions \cite{CW}, $\mathcal{F}(\mathbf{K_{1}},\mathbf{K_{2}},\omega)$ and $\mathcal{G}(\mathbf{K_{1}},\mathbf{K_{2}},\omega)$ are symmetric with respect to $\mathbf{K}_{1}$ and $\mathbf{K}_{2}$, and thus they can be Hermitian.

With these two matrices in hands and making use of the well-known trace operation of matrix together with Eq. \eqref{polarization}, the normalized CIF of the scattered field can be formulated as
\begin{small}
\begin{widetext}
\begin{align}\label{NCIF1}
 {\mathcal{C}_{n}}\left( {r{\mathbf{s}}}_{1}{,r{\mathbf{s}_{2}},\omega}%
\right)&=
\begin{cases}
\frac{\alpha_{1}\alpha_{2}(1+\mathcal{P})^2+\beta_{1}\beta_{2}(1-\mathcal{P}^2)+\varepsilon_{1}\varepsilon_{2}(1-\mathcal{P})^2}{\alpha_{1}\alpha_{2}(1+\mathcal{P})^2+(\alpha_{1}\varepsilon_{2}+\alpha_{2}\varepsilon_{1})(1-\mathcal{P}^2)+\varepsilon_{1}\varepsilon_{2}(1-\mathcal{P})^2}\frac{\Bigl|\text{Tr}\bigl[\mathcal{F^{^\top}}({{\mathbf{K}}}_{1}{,{\mathbf{K}_{2}},\omega})\cdot \mathcal{G}({{\mathbf{K}}}_{1}{,{\mathbf{K}_{2}},\omega})\bigr]\Bigr|^2}{\text{Tr}\bigl[\mathcal{F^{^\top}}({{\mathbf{K}}}_{1}{,{\mathbf{K}_{1}},\omega})\cdot \mathcal{G}({{\mathbf{K}}}_{1}{,{\mathbf{K}_{1}},\omega})\bigr]\text{Tr}\bigl[\mathcal{F^{^\top}}({{\mathbf{K}}}_{2}{,{\mathbf{K}_{2}},\omega})\cdot \mathcal{G}({{\mathbf{K}}}_{2}{,{\mathbf{K}_{2}},\omega})\bigr]}, & A_{x}\geq A_{y}\\
\\
\frac{\alpha_{1}\alpha_{2}(1-\mathcal{P})^2+\beta_{1}\beta_{2}(1-\mathcal{P}^2)+\varepsilon_{1}\varepsilon_{2}(1+\mathcal{P})^2}{\alpha_{1}\alpha_{2}(1-\mathcal{P})^2+(\alpha_{1}\varepsilon_{2}+\alpha_{2}\varepsilon_{1})(1-\mathcal{P}^2)+\varepsilon_{1}\varepsilon_{2}(1+\mathcal{P})^2} \frac{\Bigl|\text{Tr}\bigl[\mathcal{F^{^\top}}({{\mathbf{K}}}_{1}{,{\mathbf{K}_{2}},\omega})\cdot \mathcal{G}({{\mathbf{K}}}_{1}{,{\mathbf{K}_{2}},\omega})\bigr]\Bigr|^2}{\text{Tr}\bigl[\mathcal{F^{^\top}}({{\mathbf{K}}}_{1}{,{\mathbf{K}_{1}},\omega})\cdot \mathcal{G}({{\mathbf{K}}}_{1}{,{\mathbf{K}_{1}},\omega})\bigr]\text{Tr}\bigl[\mathcal{F^{^\top}}({{\mathbf{K}}}_{2}{,{\mathbf{K}_{2}},\omega})\cdot \mathcal{G}({{\mathbf{K}}}_{2}{,{\mathbf{K}_{2}},\omega})\bigr]},  & A_{x}\leq A_{y}
\end{cases}
\end{align}
\end{widetext}
\end{small}
where $\top$ represents transpose operation. 

Eq. \eqref{NCIF1} is one of the main results in the present paper. It builds a closed-form relation that associates the normalized CIF with the PPM and the PSM as well as the spectral degree of polarization $\mathcal{P}$ of the incident field. The first line in Eq. \eqref{NCIF1} is suitable for the case where the incoming source is polarized along the $x$ axis, and the second line is applied to the situation where the incoming source is polarized along the $y$ axis. The transition from the first line to the second line will occur only when $\mathcal{P}$ is replaced by $-\mathcal{P}$, therefore, the normalized CIF of the scattered field produced by an electromagnetic wave polarized along $y$ axis can be easily recovered from that of the scattered field produced by an electromagnetic wave polarized along $x$ axis, and vice versa. The fact that the normalized CIF of the scattered field split into two parts, one having to do with $\mathcal{P}$ and the other with PPM and PSM, is not accidental but we have assumed that the scattering is weak (in the sense of the first-order Born approximation). The first part shows that the dependence of the normalized CIF on $\mathcal{P}$ is only determined by the scattering polar angle $\theta$ and scattering azimuth $\phi$. As we will see, depending on $\theta$ and $\phi$, the normalized CIF itself can be monotonically increasing or nonmonotonic with $\mathcal{P}$ in the region $[0,1]$. The second part shows that the normalized CIF is intimately related to the trace of the product of the PSM and the transpose of the PPM. For a special case where the spatial distributions of the scattering potentials of particles of different types are similar and the same is true of their density distributions, i.e., $\widetilde{C}_{f_{pq}}(-\mathbf{K},\mathbf{K},\omega)\approx\widetilde{C}_{f}(-\mathbf{K},\mathbf{K},\omega)$ and $\widetilde{C}_{g_{pq}}(-\mathbf{K},\mathbf{K},\omega)\approx\widetilde{C}_{g}(-\mathbf{K},\mathbf{K},\omega)$, Eq. ($\ref{NCIF1}$) can have a simpler form, viz., 
\begin{widetext}
\begin{align}\label{NCIF2}
{\mathcal{C}_{n}}\left( {r{\mathbf{s}}}_{1}{,r{\mathbf{s}_{2}},\omega}%
\right)&=
\begin{cases}
\frac{1}{L^4}\frac{\alpha_{1}\alpha_{2}(1+\mathcal{P})^2+\beta_{1}\beta_{2}(1-\mathcal{P}^2)+\varepsilon_{1}\varepsilon_{2}(1-\mathcal{P})^2}{\alpha_{1}\alpha_{2}(1+\mathcal{P})^2+(\alpha_{1}\varepsilon_{2}+\alpha_{2}\varepsilon_{1})(1-\mathcal{P}^2)+\varepsilon_{1}\varepsilon_{2}(1-\mathcal{P})^2}\Bigl|\text{Tr}\bigl[\mathcal{U^{^\top}}({{\mathbf{K}}}_{1}{,{\mathbf{K}_{2}},\omega})\cdot \mathcal{G}({{\mathbf{K}}}_{1}{,{\mathbf{K}_{2}},\omega})\bigr]\Bigr|^2,   & A_{x}\geq A_{y}\\
\\
\frac{1}{L^4}\frac{\alpha_{1}\alpha_{2}(1-\mathcal{P})^2+\beta_{1}\beta_{2}(1-\mathcal{P}^2)+\varepsilon_{1}\varepsilon_{2}(1+\mathcal{P})^2}{\alpha_{1}\alpha_{2}(1-\mathcal{P})^2+(\alpha_{1}\varepsilon_{2}+\alpha_{2}\varepsilon_{1})(1-\mathcal{P}^2)+\varepsilon_{1}\varepsilon_{2}(1+\mathcal{P})^2}\Bigl|\text{Tr}\bigl[\mathcal{U^{^\top}}({{\mathbf{K}}}_{1}{,{\mathbf{K}_{2}},\omega})\cdot \mathcal{G}({{\mathbf{K}}}_{1}{,{\mathbf{K}_{2}},\omega})\bigr]\Bigr|^2,  & A_{x}\leq A_{y}
\end{cases}
\end{align}
\end{widetext}
where
\begin{align}\label{matrix22}
    \mathcal{U}(\mathbf{K_{1}},\mathbf{K_{2}},\omega)&=\Bigl[\mathcal{U}_{pq}(\mathbf{K_{1}},\mathbf{K_{2}},\omega)\Bigr]_{\mathcal{L}\times\mathcal{L}}
\end{align}
with
\begin{align}
    \mathcal{U}_{pq}(\mathbf{K_{1}},\mathbf{K_{2}},\omega)=\frac{\widetilde{C}_{f_{pq}}(-\mathbf{K}_{1},\mathbf{K}_{2},\omega)}{\sqrt{\widetilde{C}_{f}(-\mathbf{K}_{1},\mathbf{K}_{1},\omega)}\sqrt{\widetilde{C}_{f}(-\mathbf{K}_{2},\mathbf{K}_{2},\omega)}},
\end{align}
and
\begin{align}\label{matrix4}
    \mathcal{K}(\mathbf{K_{1}},\mathbf{K_{2}},\omega)&=\Bigl[\mathcal{K}_{pq}(\mathbf{K_{1}},\mathbf{K_{2}},\omega)\Bigr]_{\mathcal{L}\times\mathcal{L}}
\end{align}
with
\begin{align}
    \mathcal{K}_{pq}(\mathbf{K_{1}},\mathbf{K_{2}},\omega)=\frac{\widetilde{C}_{g_{pq}}(-\mathbf{K}_{1},\mathbf{K}_{2},\omega)}{\sqrt{\widetilde{C}_{g}(-\mathbf{K}_{1},\mathbf{K}_{1},\omega)}\sqrt{\widetilde{C}_{g}(-\mathbf{K}_{2},\mathbf{K}_{2},\omega)}}.
\end{align}

In comparison with Eq. (\ref{NCIF1}), Eq. (\ref{NCIF2}) shows that the number of species of particles in this special case appears as a scaled factor to ensure the normalization of the CIF, and two new matrices appear, i.e., $\mathcal{U}(\mathbf{K_{1}},\mathbf{K_{2}},\omega)$ and $\mathcal{K}(\mathbf{K_{1}},\mathbf{K_{2}},\omega)$. The elements $\mathcal{U}_{pq}(\mathbf{K_{1}},\mathbf{K_{2}},\omega)$ of the former quantify the degree of angular cross-correlation of the scattering potentials of particles of different types (if $p\neq q$) or the degree of angular self-correlation of the scattering potentials of particles of the same type (if $p=q$), and the elements $\mathcal{K}_{pq}(\mathbf{K_{1}},\mathbf{K_{2}},\omega)$ of the latter quantify the degree of angular cross-correlation of the density distributions of particles of different types (if $p\neq q$) or the degree of angular self-correlation of the density distributions of particles of the same type (if $p=q$). The normalized CIF in this special case is related to the trace of $\mathcal{U}^{^\top}(\mathbf{K_{1}},\mathbf{K_{2}},\omega)\cdot\mathcal{Q}(\mathbf{K_{1}},\mathbf{K_{2}},\omega)$.

\section{numerical examples}
In the following, we will take two special hybrid particulate systems as examples to illustrate the influence of the off-diagonal elements of the PPM and the PSM on the normalized CIF of electromagnetic light waves on scattering from these two systems and the dependence of the normalized CIF itself on the spectral degree of polarization $\mathcal{P}$ of the incident field. 

(i) Firstly, we consider a system of random particles with determinate density distributions. A representative example of this model is a collection of particles suspended in a atmospheric mass, where irregular fluctuations in temperature and pressure of atmosphere turbulence usually lead the refractive indices of particles in different locations to be different and to be random functions of position space. If these particles move slowly, at least there will be no appreciable changes in their locations during the whole scattering process, and thus their density distributions may be determinate in space (\cite{Wolf2}, Sec. 6.3.1). For simplicity, we consider the situation where only two types of particles are contained in the collection, and assume that both the self-correlation functions of the scattering potentials of particles of same type and the cross-correlation functions of different types have Guassian forms, i.e., 
\begin{align}\label{distribution}
{C_{f_{pq}}}\left({{{\mathbf{{r_{1}^{\prime}}}}},{{\mathbf{{r_{2}^{\prime}}}}}%
,\omega}\right)&=A_{0}\exp{\Bigl[-\frac{\mathbf{r}_{1}^{\prime^{2}}+\mathbf{r}_{2}^{\prime^{2}}}{4\sigma_{pq}^{2}}\Bigr]}\exp{\Bigl[-\frac{(\mathbf{r}_{1}^{\prime}-\mathbf{r}_{2}^{\prime})^2}{2\eta_{pq}^{2}}\Bigr]}, \notag \\ &\qquad (p,q=1,2)
\end{align}
where $A_{0}$ is a positive real constant, and $\sigma_{pq}$ stands for the effective widths of the distribution functions of particles of the same type (if $p=q$) or of different types (if $p\neq q$), and $\eta_{pq}$ denotes the effective correlation widths of the distribution functions of particles of the same type (if $p=q$) or of different types (if $p\neq q$). 

The self-correlation functions of the density distributions of particles of the same type and the cross-correlation functions of the density distributions of particles of different types have the following forms
\begin{equation}\label{densitydistribution}
{C_{g_{pq}}}\left({{{\mathbf{{r_{1}^{\prime}}}}},{{\mathbf{{r_{2}^{\prime}}}}}%
}\right)=\sum_{m(p)}\delta^{*}(\mathbf{r}_{1}^{\prime}-\mathbf{r}_{pm})\sum_{m(q)}\delta(\mathbf{r}_{2}^{\prime}-\mathbf{r}_{qm}).
\end{equation}

In this case, from Eqs. \eqref{1} and \eqref{2}, the elements of the matrices $\mathcal{F}(\mathbf{K_{1}},\mathbf{K_{2}},\omega)$ and $\mathcal{G}(\mathbf{K_{1}},\mathbf{K_{2}},\omega)$ can be readily calculated as
\begin{align}\label{elements}
    \widetilde{C}_{f_{pq}}(-\mathbf{K}_{1},\mathbf{K}_{2},\omega)&=A_{0}\frac{2^{6}\pi^{3}\sigma_{pq}^{6}\eta_{pq}^{3}}{(4\sigma_{pq}^{2}+\eta_{pq}^{2})^{3/2}}\notag \\ & \times
    \exp{\Bigl[-\frac{\sigma_{pq}^{2}}{2}\bigl(\mathbf{K}_{1}-\mathbf{K}_{2}\bigr)^2}\Bigr]\notag \\ &\times\exp{\Bigl[-\frac{\sigma_{pq}^{2}\eta_{pq}^{2}}{2(4\sigma_{pq}^{2}+\eta_{pq}^{2})}\bigl(\mathbf{K}_{1}+\mathbf{K}_{2}\bigr)^2}\Bigr]
\end{align}
and
\begin{align}\label{elements1}
    \widetilde{C}_{g_{pq}}(-\mathbf{K}_{1},\mathbf{K}_{2},\omega)&=\sum_{m(p)}\exp{i\mathbf{K}_{1}\cdot \mathbf{r}_{pm}}\notag \\ &\times\sum_{m(q)}\exp{-i\mathbf{K}_{2}\cdot \mathbf{r}_{qm}}.
\end{align}
Once these matrix elements are known, the normalized CIF of the scattered field is straightforward from Eq. (\ref{NCIF1}).

\begin{figure}[bthp]
\centering
\includegraphics[width=8cm]{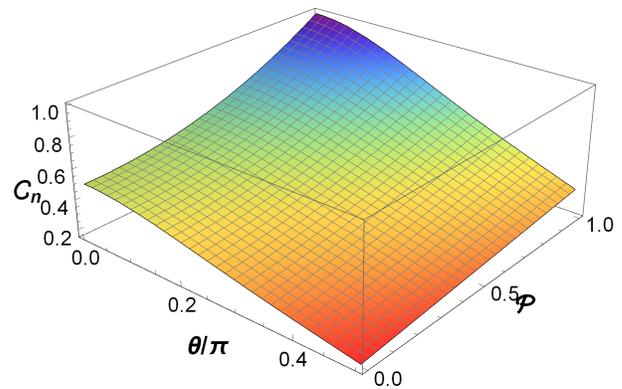}
\caption{Surface plot of the normalized CIF of electromagnetic light waves polarized along $x$ axis on scattering from a collection of random particles with determinate density distributions, as a function of the spectral degree of polarization $\mathcal{P}$ of the incident field and the dimensionless scattering polar angle $\theta/\pi$. The coordinates for the first kind of particles are set to be $(0,0.1\lambda,0)$ and $(0,-0.1\lambda,0)$, and $(0,0.2\lambda,0)$ and $(0,-0.2\lambda,0)$ for the second kind of particles. $\mathbf{s}_{0}=(0,0,1)$, $\mathbf{s}_{1}=(\sin{\theta_{1}\sin{\phi_{1}}},\sin{\theta_{1}\cos{\phi_{1}}},\cos{\theta_{1}})$,  $\mathbf{s}_{2}=(\sin{\theta\sin{\phi}},\sin{\theta\cos{\phi}},\cos{\theta})$. The parameters for calculations are $\phi=\pi/2$, $\theta_{1}=0$, $\phi_{1}=\pi/2$, $\lambda=0.6328\mu m$, $\protect\sigma _{11}=\protect\sigma _{22}=0.14\lambda$, $\protect\sigma _{12}=\protect\sigma _{21}=0.13\lambda$, $\protect\eta _{11}=\protect\eta _{22}=0.016\lambda$, $\protect\eta _{12}=\protect\eta _{21}=0.048\lambda$.}
\label{Fig 2}
\end{figure}
\begin{figure}[bthp]
\centering
\includegraphics[width=8cm]{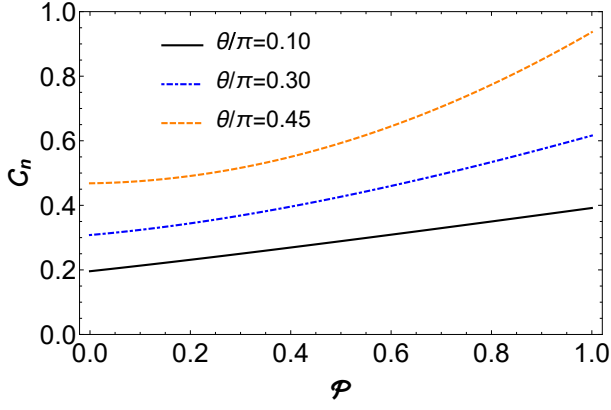}
\caption{Behaviors of the normalized CIF of the scattered field for three selective values of $\theta/\pi$. The parameters for calculations are the same as Fig. \ref{Fig 2}.}
\label{Fig 3}
\end{figure}

Fig. \ref{Fig 2} displays the surface plot of the normalized CIF of electromagnetic light waves polarized along $x$ axis on scattering from a collection of random particles with determinate density distributions, as a function of the spectral degree of polarization $\mathcal{P}$ of the incident field and the dimensionless scattering polar angle $\theta/\pi$. As shown in Fig. \ref{Fig 2}, the normalized CIF of the scattered field is a monotonically increasing function of $\mathcal{P}$ in the region $[0, 1]$ within the whole scattering polar angle, which means that the more polarized the incident field, the more intense the CIF of the scattered field. What's more, the rate (i.e., the slope $\partial\mathcal{C}_{n}/\partial\mathcal{P}$) at which the normalized CIF grows with $\mathcal{P}$ depends largely on $\theta/\pi$. To see this well, the normalized CIF of the scattered field for three selective values of $\theta/\pi$ is separately plotted in Fig. \ref{Fig 3}. It is shown that the dependence of the normalized CIF on $\mathcal{P}$ transits gradually from nonlinear to linear with the increase of $\theta/\pi$. The appearance of such a linear relation between the normalized CIF and $\mathcal{P}$ in the large scattering polar angle results from the fact that the slope $\partial\mathcal{C}_{n}/\partial\mathcal{P}$ in the neighborhood of $\theta/\pi=0.5$ is independent of $\mathcal{P}$, i.e., $\frac{\partial\mathcal{C}_{n}}{\partial\mathcal{P}}\bigl|_{\theta/\pi\approx 0.5}\propto\frac{1}{2}$, which can be readily verified from Eq. \eqref{CIF}.
\begin{figure}[bthp]
\centering
\includegraphics[width=8cm]{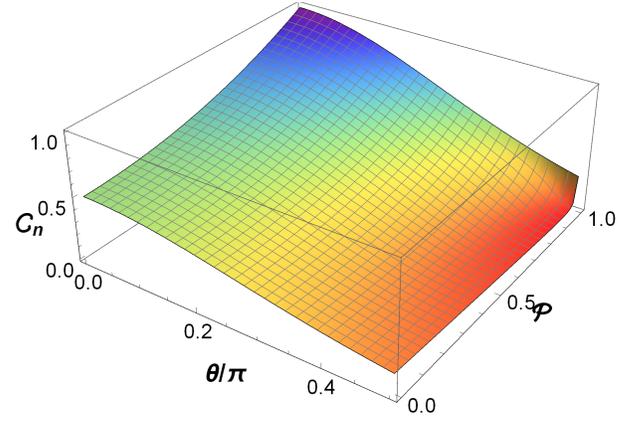}
\caption{Surface plot of the normalized CIF of electromagnetic light waves polarized along $x$ axis on scattering from a collection of random particles with determinate density distributions, as a function of $\mathcal{P}$ and $\theta/\pi$. $\phi=\pi/60$, the other parameters for calculations are the same as Fig. \ref{Fig 2}.}
\label{Fig 4}
\end{figure}
\begin{figure}[bthp]
\centering
\includegraphics[width=8cm]{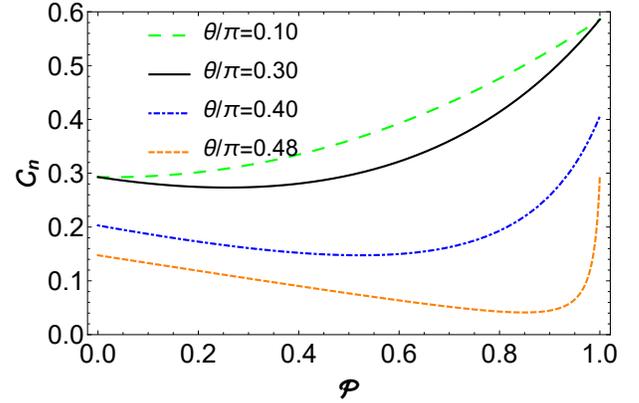}
\caption{Behaviors of the normalized CIF of the scattered field for four selective values of $\theta/\pi$. The parameters for calculations are the same as Fig. \ref{Fig 4}.}
\label{Fig 5}
\end{figure}

However, $\mathcal{C}_{n}$ won't be always monotonically increasing with $\mathcal{P}$ in the region $[0, 1]$ within the whole scattering polar angle, when the scattering azimuth angle $\phi$ happens to change. The surface plot of the normalized CIF of the scattered field in the case of $\phi=\pi/60$ is plotted in Fig. \ref{Fig 4}, as a function of $\mathcal{P}$ and $\theta/\pi$. One may see that only within very small scattering polar angles can $\mathcal{C}_{n}$ be monotonic increasing functions of $\mathcal{P}$, otherwise will it be non-monotonic with $\mathcal{P}$ in the region $[0, 1]$, which means that the claim in the case of $\phi=\pi/2$, i.e., the more polarized the incident field, the more intense the CIF of the scattered field, won't continuously hold. To see this more clearly, the normalized CIF of the scattered field for four selective values of $\theta/\pi$ is also separately plotted in Fig. \ref{Fig 5}. It is shown that, within large polar scattering angles, the normalized CIF has a manifestation of decreasing first and then increasing gradually with the growth of $\mathcal{P}$, and the decay range of $\mathcal{C}_{n}$ at the $\mathcal{P}$ axis will be greatly extended as $\theta/\pi$ raises. This reveals a truth that even if the less polarized incident field can also produce the highly intense CIF, which is quite different from the result for the incident field polarized along the $x$ axis in \cite{Li}. When $\theta/\pi$ becomes large enough, for example, $\theta/\pi=0.48$, $\mathcal{C}_{n}$ decays first in an obvious linear fashion and then grows rapidly in a nonlinear fashion within the neighborhood of $\mathcal{P}=1$. Such a dramatic change happening to $\mathcal{C}_{n}$ can have a well-defined explanation from its slope, which can be approximately computed as $\frac{\partial\mathcal{C}_{n}}{\partial\mathcal{P}}\bigl|_{\mathcal{P}\approx 1}\propto\frac{1}{1+\cos{2\theta}}$. The denominator $1+\cos{2\theta}$ will tend to zero with the approach of $\theta/\pi$ to $0.5$, naturally leading the growth of $\mathcal{C}_{n}$ with $\mathcal{P}$ at this region to be very dramatic. This clearly implies that, for the highly polarized incident field, the changes of the normalized CIF of the scattered field in large scattering polar angles are much more sensitive to $\mathcal{P}$ than that in small scattering polar angles. 


\begin{figure}[bthp]
\centering
\includegraphics[width=8cm]{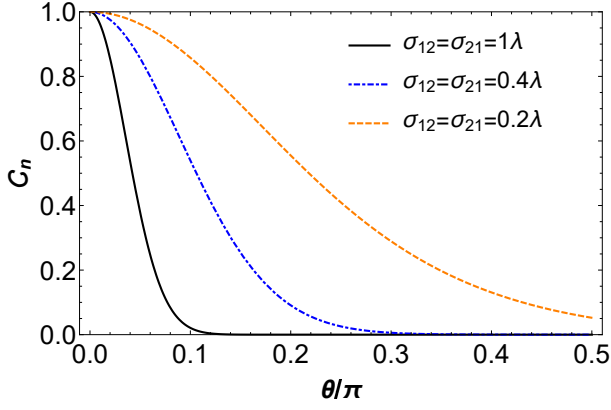}
\caption{Plots of the influence of the effective width $\sigma_{12}$ on the normalized CIF of the scattered field. The parameters for calculations are $\phi=\pi/2$, $\theta_{1}=0$, $\phi_{1}=\pi/2$, $\mathcal{P}=1$, $\protect\sigma _{11}=\protect\sigma _{22}=0.1\lambda$, $\protect\eta _{11}=\protect\eta _{22}=0.01\lambda$, $\protect\eta _{12}=\protect\eta _{21}=0.03\lambda$.}
\label{Fig 6}
\end{figure}
\begin{figure}[bthp]
\centering
\includegraphics[width=8cm]{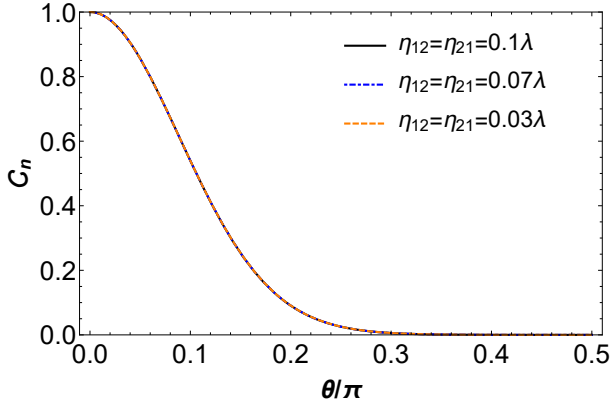}
\caption{Plots of the influence of effective correlation width $\eta_{12}$ on the normalized CIF of the scattered field. $\protect\sigma _{12}=\protect\sigma _{21}=0.4\lambda$, and the other parameters for calculations are the same as Fig. \ref{Fig 6}.}
\label{Fig 7}
\end{figure}

Now let us turn our attention to the effects of the off-diagonal elements of the PPM on the normalized CIF of the scattered field. Fig. \ref{Fig 6} depicts the behaviors of the normalized CIF of the scattered field for different effective widths $\sigma_{12}$. It is found that when the effective width $\sigma_{12}$ decreases, the normalized CIF of the scattered field can be improved greatly, which means that even if the cross-correlations between the scattering potentials of  particles of different types are weak, they can still effect the normalized CIF strongly. In comparison to the effective width $\sigma_{12}$, the effective correlation width $\eta_{12}$ can have a negligible influence on the normalized CIF of the scattered field, as can be seen from Fig. \ref{Fig 6}. This is because the effective width $\sigma_{12}$ and the effective correlation width $\eta_{12}$ meet the relation $\sigma_{12}/\eta_{12}\gg 1$ in our numerical calculations. In this case, the Gaussian Schell-model distributions in Eq. \eqref{distribution} can reduce to quasihomogeneous distributions, in which the well-known reciprocity relation holds \cite{DVW, TD}, leading the effective correlation width $\eta_{12}$ to have no consequence on the normalized CIF of the scattered field.

(ii) We now restrict our attention to the second model, i.e., a system of determinate particles with random density distributions. A representative example is a collection of cells suspended in a solution, where the refractive indices of different cells are well-defined functions in space, but their density distributions in the solution may be random (\cite{Wolf2}, Sec. 6.3.1). The current model has been taken into account preliminarily in \cite{Tong}, where use has been made of the assumption that the density distributions of particles of different types are similarly distributed in space. Here we will relax this constraint, and 
concern ourselves with a more general case, viz.,
\begin{align}\label{distribution2}
{C_{g_{pq}}}\left({{{\mathbf{{r_{1}^{\prime}}}}},{{\mathbf{{r_{2}^{\prime}}}}}}\right)&=C_{0}\exp{\Bigl[-\frac{\mathbf{r}_{1}^{\prime^{2}}+\mathbf{r}_{2}^{\prime^{2}}}{4\gamma_{pq}^{2}}\Bigr]}\exp{\Bigl[-\frac{(\mathbf{r}_{1}^{\prime}-\mathbf{r}_{2}^{\prime})^2}{2\delta_{pq}^{2}}\Bigr]}, \notag \\ &\qquad (p,q=1,2)
\end{align}
where $C_{0}$ is also a positive real constant, and $\gamma_{pq}$ and $\delta_{pq}$ have the same meaning as $\sigma_{pq}$ and $\eta_{pq}$, respectively. 

The self-correlation functions of the scattering potentials of particles of the same type and the cross-correlation functions of different types now have forms
\begin{equation}\label{densitydistribution1}
{C_{f_{pq}}}\left({{{\mathbf{{r_{1}^{\prime}}}}},{{\mathbf{{r_{2}^{\prime}}}}}%
,\omega}\right)= f_{p}^{*}(\mathbf{r_{1}}^{\prime},\omega)f_{q}(\mathbf{r_{2}}^{\prime},\omega).
\end{equation}
where
\begin{equation}\label{scatteringpotential2}
f_{p}(\mathbf{r}^{\prime},\omega)=B_{0}\exp{\Bigl(-\frac{\mathbf{r}^{\prime 2}}{2\zeta_{p}^{2}}\Bigr)}
\end{equation}
is the scattering potentials of the $p$th-type particle, with $B_{0}$ being a positive real constant. 
From Eqs. \eqref{1} and \eqref{2}, it is now seen that the elements of the matrices $\mathcal{F}(\mathbf{K_{1}},\mathbf{K_{2}},\omega)$ and $\mathcal{G}(\mathbf{K_{1}},\mathbf{K_{2}},\omega)$ have become
\begin{align}\label{elements11}
    \widetilde{C}_{g_{pq}}(-\mathbf{K}_{1},\mathbf{K}_{2},\omega)&=C_{0}\frac{2^{6}\pi^{3}\gamma_{pq}^{6}\delta_{pq}^{3}}{(4\gamma_{pq}^{2}+\delta_{pq}^{2})^{3/2}}\notag \\ & \times\exp{\Bigl[-\frac{\gamma_{pq}^{2}}{2}\bigl(\mathbf{K}_{1}-\mathbf{K}_{2}\bigr)^2}\Bigr]\notag \\ &\times\exp{\Bigl[-\frac{\gamma_{pq}^{2}\delta_{pq}^{2}}{2(4\gamma_{pq}^{2}+\delta_{pq}^{2})}\bigl(\mathbf{K}_{1}+\mathbf{K}_{2}\bigr)^2}\Bigr]
\end{align}
and
\begin{align}\label{elements3}
    \widetilde{C}_{f_{pq}}(-\mathbf{K}_{1},\mathbf{K}_{2},\omega)&=B_{0}^{2}(2\pi\zeta_{p})^6\exp{\Bigl[-\frac{1}{2}\zeta_{p}^{2}\bigl(\mathbf{K}_{1}^{2}+\mathbf{K}_{2}^{2}\bigr)}\Bigr].
\end{align}
With these matrix elements in hands, the normalized CIF of the scattered field is straightforward from Eq. (\ref{CIF}) once again.

\begin{figure}[bthp]
\centering
\includegraphics[width=8cm]{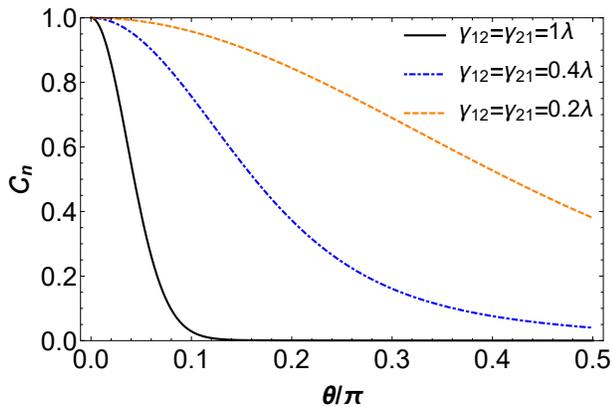}
\caption{Plots of the influence of the effective width $\gamma_{12}$ on the normalized CIF of electromagnetic light waves polarized along $x$ axis on scattering from a collection of determinate particles with random density distributions. The parameters for
calculations are $\phi=\pi/2$, $\theta_{1}=0$, $\phi_{1}=\pi/2$, $\zeta_{1}=0.2\lambda$, $\zeta_{2}=0.1\lambda$, $\mathcal{P}=1$, $\protect\gamma _{11}=\protect\gamma _{22}=0.1\lambda$, $\protect\delta _{11}=\protect\delta _{22}=0.02\lambda$, $\protect\delta _{12}=\protect\delta _{21}=0.01\lambda$.}
\label{Fig 8}
\end{figure}
\begin{figure}[bthp]
\centering
\includegraphics[width=8cm]{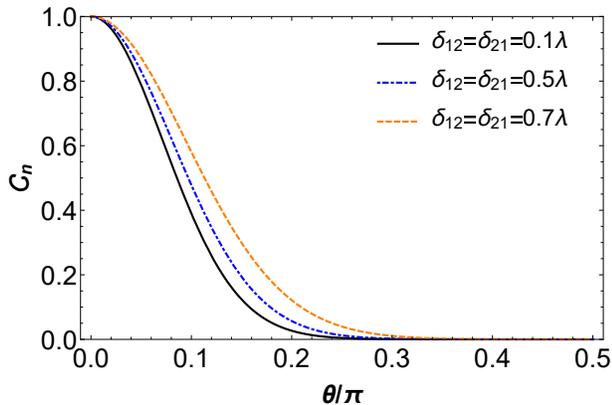}
\caption{Plots of the influence of the effective correlation width $\gamma_{12}$ on the normalized CIF of the scattered field. $\protect\gamma _{11}=\protect\gamma _{22}=0.1\lambda$, $\protect\gamma _{12}=\protect\gamma _{21}=1\lambda$, $\protect\delta _{11}=\protect\delta _{22}=0.2\lambda$. The other parameters for calculations are the same as Fig. \ref{Fig 8}.}
\label{Fig 9}
\end{figure}

We now concern ourselves with the effects of the off-diagonal elements of the PSM on the normalized CIF of electromagnetic light waves polarized along $x$ axis on scattering from a collection of determinate particles with random density distributions. Fig. \ref{Fig 8} presents the behaviors of the normalized CIF of the scattered field for three effective widths $\gamma_{12}$. It is found that when the effective width $\eta_{12}$ decreases, the normalized CIF of the scattered field can also be improved significantly, and the normalized CIF in the current model can still has an appreciable value even if the scattered field is observed at the large scattering polar angle, compared to the first model. Fig. \ref{Fig 9} shows the distributions of the normalized CIF for three different effective correlation widths $\delta_{12}$, where $\delta_{12}$ now is comparable to $\gamma_{12}$. From Fig. \ref{Fig 9} it follows that the effective angular width of the normalized CIF increases with the increase of the cross-correlation width of the correlation function in Eq. \eqref{distribution2}, which demonstrates intuitively that the influence of the effective correlation width $\delta_{12}$ on the normalized CIF of the scattered field can be observed provided that the constraint $\gamma_{12}/\delta_{12}\gg 1$ is relieved. 

\section{Summary and Discussion}
In summary, we have developed a theoretical framework in the spherical polar coordinate system to systematically treat the CIF of electromagnetic light waves on scattering from a collection of particles of $\mathcal{L}$ types. Two $\mathcal{L}\times\mathcal{L}$ matrices called PPM and PSM were introduced to jointly formulate the normalized CIF of the scattered field. These two matrices not only provide sufficient amount of information to determine all the second-order statistical properties of the scattered field produced on scattering from a hybrid particulate system, but also can largely simplify the theoretical procedures of the scattering of light from complex collection of scatters. We derived a closed-form relation that associates the normalized CIF with the PPM and the PSM as well as the spectral degree of polarization $\mathcal{P}$ of the incident field, showing that the normalized CIF is intimately related to the trace of the product of the PSM and the transpose of the PPM. For a special case where the spatial distributions of scattering potentials of particles of different types are similar and the same is true of their density distributions, these two matrices can reduce to another two new matrices whose elements separately quantify the degree of angular correlation of the scattering potentials of particles and their density distributions, and the number of species of particles in this special case appears as a scaled factor to ensure the normalization of the CIF. Two special hybrid particulate systems were taken as examples to illustrate the effects of the off-diagonal elements of the PPM and the PSM on the normalized CIF of the scattered field, and the dependence of the normalized CIF on $\mathcal{P}$ was also discussed in detail. We found that the non-zero cross-correlation between particles of different types in the collection can affect strongly on the normalized CIF of the scattered field, and the normalized CIF itself can either be monotonically increasing or be non-monotonic with $\mathcal{P}$ in the region $[0,1]$, depending on $\theta$ and $\phi$. 

Although there have been many studies on the influence of polarization property of the incident field on the CIF of the scattered fields generated by different scattering media \cite{Li, ZZ, LCC, LCC1, Ding, Wang}, but we still would like to emphasize the difference between our results and theirs. Firstly, the previous efforts to explore the CIF of the scattered field were mainly confined within the so-called scattering plane, that is, the effects of the scattering azimuth angle $\phi$ on the CIF were usually not taken into account. However, as we have seen from Fig. \ref{Fig 3} and Fig. \ref{Fig 5} that $\phi$ is a nontrivial element that determines the dependence of $\mathcal{C}_{n}$ on $\mathcal{P}$. Secondly, Eq. \eqref{NCIF1} is the final expression to show how the normalized CIF of electromagnetic scattered fields changes, depending on $\mathcal{P}$. This qualitative and quantitative relation between the normalized CIF and $\mathcal{P}$ hasn't been noticed before. In the previous works, the normalized CIF was always shown to be a function of polarization amplitudes $A_{x}$ and $A_{y}$ just like Eq. \eqref{NCIF11} here, and even to be a function of $\frac{A_{y}^{2}}{A_{x}^{2}+A_{y}^{2}}$ (please see Fig. 4 in \cite{Li}), which is still not the spectral degree of polarization $\mathcal{P}$. Unlike the results in \cite{Li}, the normalized CIF as a function of $\mathcal{P}$ can be nonmonotonic with $\mathcal{P}$ in the range $[0, 1]$ even if electromagnetic light waves are polarized along the $x$ axis. Additionally, when $\theta_{1}=\theta=0$, $\mathcal{C}_{n}$ won't remain constant no matter what $\mathcal{P}$ the incident field carries. These elaborate analyses on the relation between the normalized CIF and $\mathcal{P}$ in this work may have potential applications in the areas of ghost scattering \cite{Cheng} and ghost imaging \cite{TCK, TKS}. Our results may be more conducive to retrieve a high quality of ghost image via collecting intensity correlation information at proper observation angles, when an electromagnetic light wave with a certain $\mathcal{P}$ value is used as incoming source. 


\section{Acknowledgements}
The author acknowledges Zhenfei Jiang for her passionate assistance during the author visit to University of Rochester. Financial support was provided by Fundamental Research Funds for the Central Universities No. 2682022CX040.

\end{document}